\documentstyle[aps,multicol,prb,epsf]{revtex}
\begin{document}
\draft
\title{Quantum dots with two electrons: Singlet-triplet transitions}
\author{O. Entin-Wohlman\cite{OEW} }
\address{Center of Advanced Studies,
Norwegian Academy of Sciences, Oslo 0271, Norway}
\author{Amnon Aharony}
\address{\it School of Physics and Astronomy, Raymond and Beverly Sackler Faculty
of Exact Sciences, \\ Tel Aviv University, Tel Aviv 69978, Israel}
\author{and Y. Levinson}
\address{\it Department of Condensed Matter Physics, The Weizmann Institute of Science, Rehovot 76100, Israel}

\date{\today}
\maketitle

\begin{abstract}
The magnetic character of the ground-state of two electrons on a
double quantum dot, connected in series to left and right
single-channel leads, is considered. By solving exactly for the
spectrum of the two interacting electrons, it is found that the
coupling to the continuum of propagating states on the leads, in
conjunction with the electron-electron interactions, may result in
a delocalization of the bound state of the two electrons. This, in
turn, reduces significantly the range of the Coulomb interaction
parameters over which singlet-triplet transitions can be realized.
It is also found that the coupling to the leads favors the singlet
ground-state.
\end{abstract}

\begin{multicols}{2}

\section{Introduction}

The  spin state of the many electron ground-state is determined by
the interplay between the kinetic and the electrostatic
interactions. According to Hund's law the ground-state of
electrons in a partially filled shell of an atom has the maximal
possible spin, in order to minimize the electrostatic repulsion.
On the other hand,  Anderson's super-exchange antiferromagnetic
interaction, which favors zero total spin, arises from the
reduction in the ground-state energy brought about by the hopping
of electrons between adjacent ions. Another example is realized in
gases made of homonuclear diatomic molecules, in which the
distance between the nuclei determines the magnetic properties of
the gas; in other words, the ground-state of the molecule can be
either a triplet or a singlet.

The possibility to study these rules as function of controlled
parameters, and in particular to observe deviations from them, has
been opened up in recent years with the intensive experimental and
theoretical investigations of small confined systems. Devices
based on quantum dots formed in GaAs heterostructures allow to
probe the electronic states in situations where there are only a
few electrons in the system, as function of, e.g., the number of
electrons in the sample, the spacing of the single-electron energy
levels, the internal structure of the dots (e. g. the distance
and the coupling between internal parts)
or the coupling of the system to external leads. Thus,
modulating the single-particle spectrum by  a magnetic field,
applied perpendicularly to the plane of the electron gas,
\cite{tarucha1,tarucha2} has produced a structure of the
conductance peaks that has been interpreted in terms of
singlet-triplet transitions of the last two electrons in the dot
\cite{pustilnik1,pustilnik2} (see also Ref. \onlinecite{eto}).
This has been observed on vertical quantum dots, \cite {tarucha2}
and also in lateral ones. \cite{schmid} The ground-state spin of
{\it chaotic} quantum dots has been studied by tracing the
conductance peak spacing as function of a weak parallel magnetic
field (which couples primarily to the spins). \cite{folk} A
numerical investigation of such a configuration analyzed the
influence of the exchange interaction on the peak structure.
\cite{berkovits} In a similar fashion, the singlet-triplet
transitions in such dots have been attriubted to avoided-crossings
in the many-electron states, and the relation between those and
kinks in the conductance pattern has been explored.
\cite{baranger}

When the device consists of two or more coupled dots, another
controllable parameter comes into the play: the interdot coupling.
Experimentally, \cite{lee} it has been found that this coupling
shows up in the conductance peak positions. Theoretically, the
effect of the interdot distance on the states of a few electrons
confined in a parabolic potential has been analyzed using
mean-field methods, \cite{palacios} the Kohn-Sham equation,
\cite{partoens} and numerical diagonalization.\cite{martin}
Different spin states of double-quantum-dots systems have been
also studied using the numerical renormalization group method.
\cite{izumida}

Here we present an exact analytical solution for the eigenvalues
and eigenfunctions of two electrons on a double-dot system, which
is coupled to two single-channel leads. The two electrons interact
while they are on the dots, and we include in the calculation the
direct Coulomb and the exchange interactions between the dots, and
the on-site Hubbard interactions on the dots. We obtain the
condition for the singlet-triplet transitions of the ground-state
energies, and in particular examine the role of the {\it
delocalization} effect of the interactions: It has been found in
our previous works, \cite{we1,we2} as well as in other studies,
\cite{vidal,weinmann} that the interplay between the hybridization of
the localized single-particle states on the dot with the
propagating states on the leads, and the electron-electron
interactions, may promote one of the electrons to the continuum.
If the system had only one doubly bound state, then the above
`promotion' results in a `delocalization' of the ground state.
Once delocalized,
the singlet and the triplet states become
degenerate.

Our earlier work included only the case with on-site Coulomb repulsion.
In that case, the bound ground state is a singlet. As the coupling of the
dots with the leads was increased, the parameter range where this bound state
exists was decreased, and the delocalization was accompanied by a transition
from the singlet to a degenerate singlet-triplet `metallic' state.
In the present paper we add the effects of direct and exchange interactions.
In the absence of the coupling to the leads, it is well known \cite{harrison}
that the ground state changes from a singlet to a triplet upon
increasing the exchange interaction. Here we investigate what happens to that
singlet-triplet transition in the presence of the coupling to the leads.

After discussing our model Hamiltonian in Sec. II, we present the exact
two-electron solution in Sec. III. Section IV then discusses the
singlet-triplet transition, mainly for a special choice of the parameters
where it is easiest to explore the solution analytically. Our results are
then summarized in Sec. V.

\section{Hamiltonian}

 The Hamiltonian of the model reads
 \begin{eqnarray}
 {\cal H}={\cal H}_{\rm sp}+{\cal H}_{\rm c}.\label{ham}
 \end{eqnarray}
 The single-particle tight binding Hamiltonian is
 \begin{eqnarray}
 {\cal H}_{\rm sp}=\sum_{{\rm i}\sigma}
 \epsilon_{{\rm i}}c^{\dagger}_{{\rm i}\sigma}
 c_{{\rm i}\sigma}-\sum_{{\rm ij}\sigma}t_{\rm ij}
 c^{\dagger}_{{\rm i}\sigma}c_{{\rm j}\sigma},
 \label{hamsp}
 \end{eqnarray}
 where $c^{\dagger}_{{\rm i}\sigma}$ creates an electron with spin
$\sigma$ on site i. The site energies $\epsilon_{{\rm i}}$ are different
from zero only on the `quantum dots', which can be viewed as `impurities'
on the lattice. In what follows, we specifically consider two
neighboring quantum dots, which will be denoted by $\ell$ and $r$, and
study the symmetric case where $\epsilon_{\ell ,r}=\epsilon_{0}$.
The hopping matrix elements $t_{\rm ij}$ are divided into three kinds:
the hopping among the `dots' is denoted by $t_D$ ($\equiv t_{\ell r}$),
the hopping between a dot and a neighboring site on the `lead' is denoted
by $t_0$ (e. g. $t_{\ell {\rm i}}$ for i $\neq r$),
and the hopping between neighboring `lead' sites is denoted by
$t = 1$, setting the units of energy.

The Coulomb interactions are assumed to exist only among electrons which
sit on the dots.
Generally, this interaction has the form
\begin{eqnarray}
{\cal H}_{\rm c}=\sum_{\rm ijmn}\Gamma_{\rm ij}^{\rm mn}
\sum_{\sigma\sigma '}c^{\dagger}_{{\rm i}\sigma}c^{\dagger}_{{\rm j}\sigma '}
c_{{\rm m}\sigma '}c_{{\rm n}\sigma},
\end{eqnarray}
where
\begin{eqnarray}
\Gamma_{{\rm ij}}^{\rm mn}=\int d{\bf r}\int d{\bf r}'
v({\bf r}-{\bf r}')\varphi^{\ast}_{\rm i}({\bf r})
\varphi^{\ast}_{\rm j}({\bf r}')\varphi_{\rm m}({\bf r}')
\varphi_{\rm n}({\bf r}),
\label{general}
\end{eqnarray}
while $\varphi_{\rm i}({\bf r})$ is the Wannier wave function
of an electron localized at site i.
It is usually assumed that the dominant terms will be
those in which the indices are equal pair-wise. \cite{kurland}
Here we follow the parametrization used in magnetic coupling
studies, \cite{entin} in which the case ${\rm i=j=m=n}$
is treated separately.
Neglecting the coefficient with ${\rm i}={\rm j} \neq {\rm m}={\rm n}$,
which (when negative) leads to the superconductivity vertex, we are left
with three possible parameters:
\begin{eqnarray}
2\Gamma_{\rm ii}^{\rm ii}&=&{\rm U},\nonumber\\
2\Gamma_{\rm ij}^{\rm ji}&=&{\rm V},\ \
2\Gamma_{\rm ij}^{\rm ij}={\rm K},\ \ \ {\rm i} \neq {\rm j},
\label{me}
\end{eqnarray}
where ${\rm U},{\rm V}$ and ${\rm K}$ are the (intra-dot) Hubbard,
(inter-dot) direct,
and (inter-dot) exchange interactions, respectively.
Then
\begin{eqnarray}
{\cal H}_{\rm int}&=&{\rm U}\sum_{\rm i}\hat{n}_{{\rm i}\uparrow}
\hat{n}_{{\rm i}\downarrow}\nonumber\\
&+&\frac{1}{4}(2{\rm V}-{\rm K})\sum_{{\rm i}\neq {\rm j}}
\hat{n}_{{\rm i}}\hat{n}_{{\rm j}}
-{\rm K}\sum_{{\rm i}\neq{\rm j}}{\bf S}_{{\rm i}}\cdot{\bf S}_{{\rm j}},
\label{Hwe}
\end{eqnarray}
where
\begin{eqnarray}
\hat{n}_{{\rm i}\sigma}&=&c^{\dagger}_{{\rm i}\sigma}c_{{\rm i}\sigma},\ \
\hat{n}_{{\rm i}}=\sum_{\sigma}\hat{n}_{{\rm i}\sigma},\nonumber\\
{\bf S}_{{\rm i}}&=&\sum_{\sigma\sigma '}c^{\dagger}_{{\rm i}\sigma}
{\vec \sigma}_{\sigma\sigma '}c_{{\rm i}\sigma '}
\end{eqnarray}
and ${\vec \sigma}_{\sigma\sigma '}$ is the vector of Pauli's spin matrices.
It is important
to note that with the above approximation for the Coulomb
interaction vertex, the interaction Hamiltonian becomes {\it
spin-dependent}.

A similar Hamiltonian has been used to study the effects of the spin states
on the conducting properties of confined mesoscopic quantum dots.
\cite{pustilnik1,pustilnik2,folk,berkovits}
In these studies, ${\rm i}$ represents some orbital state on the dot.
In Ref. \onlinecite{kurland} it has been argued
that when the Thouless conductance of the confined (mesoscopic)
system is large, the Coulomb vertices can be parametrized in terms
of two coupling constants, independent of the orbital indices.
In terms of our parameters, these
are  the charging energy, ${\rm E}_{c}=(2{\rm
V}-{\rm K})/4$, and the exchange
energy, ${\rm J}={\rm K}$,
while ${\rm U} ={\rm V}+{\rm K}$.
It is straightforward to apply our results to that case.

In our special case of the double dot,
the energies
$\Gamma$ in Eq. (\ref{me}) differ from zero only for ${\rm i},\ {\rm j}=
\ell,r$, and thus
 \begin{eqnarray}
 {\cal H}_{\rm int}&=&{\rm U}(\hat{n}_{\ell\uparrow}\hat{n}_{\ell\downarrow}+
 \hat{n}_{r\uparrow}\hat{n}_{r\downarrow})
 \nonumber\\
 &+&({\rm V}-\frac{{\rm K}}{2})\hat{n}_{\ell}
 \hat{n}_{r}-2{\rm K}{\bf S}_{\ell}\cdot{\bf S}_{r}.\label{hamint}
 \end{eqnarray}

\section{The two-electron 'molecule'}

We will confine ourselves to a double-dot system (a `molecule'), containing
two electrons. The system is modeled by two identical
 single-level impurities, each having an on-site single-particle
 energy level $\epsilon_{\ell}=\epsilon_r=\epsilon_{0}$.
 The two impurities are coupled to one another by the interdot
 matrix element $t_{D}$.
 When the molecule is isolated ($t_0=0$), there are three degenerate
 triplet states, of energy
 $2\epsilon_{0}+{\rm V}-{\rm K}$, and three singlet states.
 One of the latter has the energy
 $2\epsilon_{0}+{\rm U}$, while the other two energies are \cite{harrison}
 \begin{eqnarray}
 {\rm E}=2\epsilon_{0}+\frac{{\rm U}+{\rm V}+{\rm K}}{2}\pm\sqrt{4t_{D}^{2}
 +\Bigl (\frac{{\rm U}-{\rm V}-{\rm K}}{2}\Bigr )^{2}}.\label{Es}
 \end{eqnarray}
 Examination of the interaction matrix elements, Eqs. (\ref{general}) and
(\ref{me}),
 shows that ${\rm U}>{\rm V}>0$ and ${\rm U}>{\rm K}$.
 Moreover, the exchange interaction ${\rm K}$ involves
 the square of the overlap
 matrix element of the two impurities, and hence is of
 the order of $t_{D}^{2}$. Therefore
 one concludes that the lowest singlet energy is
 given by the minus sign in (\ref{Es}).
 The triplet states energy is below the lowest singlet one provided that
 \begin{eqnarray}
 2t_{D}^{2}<{\rm K}({\rm U}-{\rm V}+{\rm K}).\label{st0}
 \end{eqnarray}
 One notes that when the direct and exchange Coulomb vertices
 V and K are disregarded, the
 energy of the singlet state is lowered by the interdot
 kinetic energy, producing the
 Anderson super-exchange antiferromagnetic coupling,
 \cite {anderson}  $\sim 4t_{D}^{2}/{\rm U}$. The triplet
 state becomes the ground-state once
 the exchange energy ${\rm K}$ wins over this kinetic energy.
 \cite{brouwer,eto} Finally, with the choice of parameters
 in which ${\rm U}={\rm V}+{\rm K}$ (see above), the condition becomes
 $t_{D}^{2}<{\rm K}^2$ (see also Ref. \onlinecite{pustilnik1}).
 Both $t_D$ and K  decay exponentially with the interdot distance.
  However, Eq. (\ref{general}) indicates that K is roughly of order
$t_D^2$, and thus decays faster. Therefore, one might expect a transition
from the triplet to the singlet ground state as this distance increases.

 The main purpose of the present paper is
 to study how the criterion for the
 singlet-triplet transitions, Eq. (\ref{st0}),
 is modified when the double quantum dot is
 connected to external leads, on which
 it is assumed that there are no electron-electron
 interactions. In other words,
 we study the changes in these transitions due to
 coupling the dot to a continuum of
 propagating states.
 A similar problem was discussed in Ref. \onlinecite{chudnovskiy},
 in the framework
 of the mean-field approximation, within a single-lead geometry.  Here we
 first derive {\it exactly}
 the ground-state energy of {\it two} electrons, and then
 compare the singlet and the
 triplet ground-state energies.

\begin{figure}
 \epsfclipon
\epsfxsize = 8cm
\centerline{
\vbox{
\epsffile{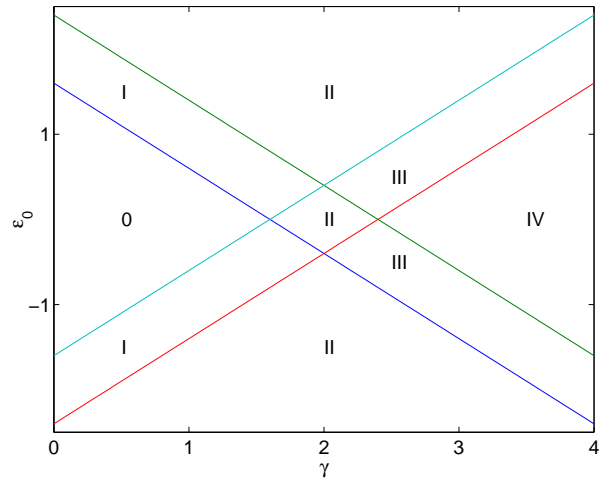}
}}
\vskip  1truecm
\caption{The single-particle bound energies. The roman numbers denote the
number of bound states. Here $t_{D}$=0.4.}
\label{fig1}
 \end{figure}

 For further simplification, we describe the external leads by single
channel one dimensional chains, with nearest neighbor hopping $t=1$.
In our previous work, \cite{we1,we2} we have shown
 that the spectrum and the wave functions of
 two interacting electrons can be obtained in terms of
 the energy spectrum and the wave functions
 of the single-particle Hamiltonian.
 For our model, cf. Eq. (\ref{hamsp}),  the latter
 can be divided into even
 and odd solutions, denoted by ``e" and ``o".
 Of particular interest here will be the regions in the
 parameter plane $\{\epsilon_{0}-\gamma\}$, where
 $\gamma =t_{0}^{2}$, where the spectrum
 has bound states, see Fig. \ref{fig1}.
 For $\epsilon_{0}<\gamma-2+t_{D}$, there appears one
 (even) bound (single-particle) state below the band
 of propagating states, of
 energy $\epsilon_{\rm B}^{\rm e}<-2$; For even smaller
 values of $\epsilon_{0}$, such that
 $\epsilon_{0}<\gamma -2 -t_{D}$ there appears a
 second (odd) bound state below the band, of
 energy $\epsilon_{\rm B}^{\rm o}$, with
 \begin{eqnarray}
 \epsilon_{\rm B}^{\rm e,o}&=&\frac{2-\gamma}{1-\gamma}
 \Bigl (\frac{\epsilon_{0}^{\rm e,o}}{2}
 \Bigr )
 +\frac{\gamma}{1-\gamma}\sqrt{\Bigl (\frac{\epsilon_{0}^{\rm e,o}}{2}
 \Bigr )^{2}+\gamma -1},
 \nonumber\\
 \epsilon_{0}^{\rm e}&=&\epsilon_{0}- t_{D},\ \ \epsilon_{0}^{\rm o}=\epsilon_{0}+ t_{D}.\label{eb}
 \end{eqnarray}
 Similarly, for $\epsilon_{0}
 >2-\gamma -t_{D}$ there appears the first bound
 state above the band, while for $\epsilon_{0}
 >2-\gamma +t_{D}$ there are two bound states above the band.
 Clearly, a necessary condition to
 have the two electrons bound in a triplet state
 is the existence of two distinct single-electron bound states,
so that each electron occupies a different `state'.
 For simplicity,
 we shall confine ourselves to the case in which
 both occur below the continuum, i.e.,
 to the lowest region II in Fig. \ref{fig1}.

 It has been shown in Refs. \onlinecite{we1}
 and \onlinecite{we2} that when
 interactions between the two electrons
 are allowed, they may lead to the
 {\it delocalization} of one (or both) of the electrons from
 the doubly bound states.  Thus, for example,
 \cite{we2} when V=K=0, in part of  region I in
 Fig. \ref{fig1} one finds that the doubly bound ground state
is replaced by a ground state in which one of the electrons is
 shifted to the band. In these parameter regions there are no
 two-electron bound states, and the system may be called `metallic'.
 Indeed, such interaction-induced
 delocalization effects have been recently observed in quantum dots containing
 two potential minima.
\cite{brodsky} This effect has also been found for two
 interacting electrons moving in a one-dimensional periodic
 structure, \cite{vidal} and for spinless fermions on strongly
 disordered chains. \cite{weinmann}
In contrast to region I,
in region II there potentially exist two doubly bound states, i. e. the
singlet and the triplet, and our goal is to study which of these states
is the ground state in the presence of this
interaction-induced delocalization.

 To solve for the spectrum of the two interacting
 electrons we proceed as follows. Let us denote the
 eigenstates and the eigenvalues of the single-particle Hamiltonian (\ref{hamsp}) by
 $\phi_{a}(n)$ and $\epsilon_{a}$, respectively,
 where $n$ is the site index.
 Consider
 first the two-electron
 states with spin $S_{z}=\pm 1$,
 \begin{eqnarray}
 |\Psi^{\sigma}\rangle &=&\sqrt{\frac{1}{2}}\sum_{ab}
 {\rm X}^{\sigma}_{ab}c^{\dagger}_{a\sigma}c^{\dagger}_{b\sigma}
 |0\rangle\nonumber\\
 &=&\sqrt{\frac{1}{2}}\sum_{ab}{\rm X}^{\sigma}_{ab}
 \sum_{nn'}\phi^{\ast}_{a}(n)\phi^{\ast}_{b}(n')
 c^{\dagger}_{n\sigma}c^{\dagger}_{n'\sigma}|0\rangle\nonumber\\
 &&\equiv \sqrt{\frac{1}{2}}\sum_{nn'}
 {\rm X}_{nn'}^{\sigma}c^{\dagger}_{n\sigma}c^{\dagger}_{n'\sigma}|0\rangle,
 \ \ \sigma =\pm 1,
 \end{eqnarray}
 where the amplitudes ${\rm X}^{\sigma}_{nn'}=-{\rm X}^{\sigma}_{n'n}$
 are antisymmetric in the
 coordinates (the origin is half way between the dots), and
 $\sum_{nn'}|{\rm X}^{\sigma}_{nn'}|^{2}=1$.
 The Schr\"{o}dinger equation for the two electrons then yields
 \begin{eqnarray}
 {\rm X}^{\sigma}_{nn'}&=&{\rm X}^{\sigma}_{r\ell}({\rm V-K})\nonumber\\
 &&\times \sum_{ab}\frac{\phi_{a}(r)\phi_{b}(\ell )-\phi_{a}(\ell )
 \phi_{b}(r)}{{\rm E}-\epsilon_{a}-\epsilon_{b}}
 \phi^{\ast}_{a}(n)\phi^{\ast}_{b}(n').
 \end{eqnarray}
 Hence, the two-particle energies, E, are given by
 the solutions of the equation
\begin{eqnarray}
 \frac{1}{\rm V-K}&=&\frac{1}{2}\sum_{ab}\frac{|\phi_{a}(\ell )
 \phi_{b}(r)-\phi_{a}(r)\phi_{b}(\ell )|^2}
 {{\rm E}-\epsilon_{a}-\epsilon_{b}}.
 \end{eqnarray}
 Making now use of the symmetry properties of $\phi_{a}$, i.e.,
 $\phi^{\rm e}_{a}(r)=\phi^{\rm e}_{a}(\ell)$, and
 $\phi^{\rm o}_{a}(r)=-\phi^{\rm o}_{a}(\ell )$,
 we see that the only contributions to the
 sum come from the cases in which $a$ is even, and $b$
 is odd, and {\it vice versa}. For the sake of brevity, in the following
expressions we replace
$|\phi^i_{a}(\ell)|=|\phi^i_{a}(r)|$ ($i=e,o$) by $|\phi^i_{a}|$.
Consequently,
 we may write the result in the form
 \begin{eqnarray}
 \frac{1}{\rm V-K}=G_{\rm eo}({\rm E}),\label{v-k}
 \end{eqnarray}
 where $G_{\rm eo}$ is a {\it noninteracting} two-electron Green's function,
 \begin{eqnarray}
 G_{\rm eo}({\rm E})&=&4\frac{|\phi^{\rm e}_{\rm B}|^{2}
 |\phi^{\rm o}_{\rm B}|^{2}}{{\rm E}-\epsilon^{\rm e}_{\rm B}
 -\epsilon^{\rm o}_{\rm B}}
 +2\sum_{\stackrel{i,j={\rm e,o}}{i\neq j}}
 \sum_{k}\frac{|\phi^{i}_{\rm B}|^{2}|\phi^{j}_{k}|^{2}}
 {{\rm E}-\epsilon^{i}_{\rm B}-\epsilon_{k}}
 \nonumber\\
 &+&\sum_{kk'}\frac{|\phi^{\rm e}_{k}|^{2}|\phi^{\rm o}_{k'}|^{2}}
 {{\rm E}-\epsilon_{k}-\epsilon_{k'}}.\label{geo}
 \end{eqnarray}
 (In writing down this equation, we have assumed the existence of two single-particle
 bound states, as mentioned above.)
 Here the subscript B denotes a single-particle bound state,
 and $k$ refers to a band state, with
 $\epsilon_{k}=-2\cos k$. In the continuum of the band energies, there is no need to distinguish between
 $\epsilon_{k}^{\rm e}$ and $\epsilon_{k}^{\rm o}$. Also,
 the sum over all states $k$ is divided into
 the sum over the even propagating states, and the sum
 over the odd ones. We give in the Appendix
 the explicit expressions for the eigenstates required for the calculation of $G$.

 Equation (\ref{v-k}) is an implicit equation for the
 exact two-electrons eigenenergies
 E. We will postpone the discussion of these solutions, and consider now
 the two-electron states with $S_{z}=0$,
 \begin{eqnarray}
 |\Psi^{0}\rangle &=&\sum_{ab}{\rm X}^{0}_{ab}
 c^{\dagger}_{a\uparrow}c^{\dagger}_{b\downarrow}
 |0\rangle
 \nonumber\\
 &=&\sum_{ab}{\rm X}^{0}_{ab}\sum_{nn'}
 \phi^{\ast}_{a}(n)\phi^{\ast}_{b}(n')c^{\dagger}_{n\uparrow}
 c^{\dagger}_{n'\downarrow}|0\rangle\nonumber\\
 &&\equiv \sum_{nn'}{\rm X}^{0}_{nn'}
 c^{\dagger}_{n\uparrow}c^{\dagger}_{n'\downarrow}|0\rangle ,
 \label{sz0}
 \end{eqnarray}
 with $\sum_{nn'}|{\rm X}^{0}_{nn'}|^{2}=1$.
 For the triplet state, the amplitudes ${\rm X}_{nn'}^{0}$
 are antisymmetric in the site indices. Then
 (inserting Eq. (\ref{sz0}) in the Schr\"{o}dinger equation)
 the energies are again given by Eqs. (\ref{v-k})
 and (\ref{geo}). For the singlet states,
 ${\rm X}^{0}_{nn'}={\rm X}^{0}_{n'n}$ and the Schr\"{o}dinger equation yields
 \begin{eqnarray}
 &&{\rm X}^{0}_{nn'}=\sum_{ab}\Biggl ({\rm U}
 \sum_{i=\ell ,r}{\rm X}^{0}_{ii}\frac{\phi_{a}(i)\phi_{b}(i)}{{\rm E}
 -\epsilon_{a}-\epsilon_{b}}
 +({\rm V+K}){\rm X}^{0}_{\ell r}\nonumber\\
 &&\times
 \frac{\phi_{a}(\ell )\phi_{b}(r)+\phi_{a}(r)
 \phi_{b}(\ell )}{{\rm E}-\epsilon_{a}-\epsilon_{b}}\Biggr )
 \phi^{\ast}_{a}(n)\phi^{\ast}_{b}(n').
 \end{eqnarray}
 We use this equation for $n,n'=\ell ,\ell$,
 $n,n'=r,r$, and $n,n'=\ell ,r$, and find two families of singlet
 solutions: (i) ${\rm X}^{0}_{\ell\ell}=-{\rm X}^{0}_{rr}$,
 ${\rm X}^{0}_{\ell r}=0$, for which
 \begin{eqnarray}
 \frac{1}{\rm U}=G_{\rm eo}({\rm E});\label{u}
 \end{eqnarray}
 (ii) ${\rm X}^{0}_{\ell\ell}={\rm X}^{0}_{rr}$, with
 \begin{eqnarray}
 1&-&({\rm U+V+K})(G_{\rm ee}({\rm E})+G_{\rm oo}({\rm E}))\nonumber\\
 &+&4{\rm U(V+K)}G_{\rm ee}({\rm E})G_{\rm oo}({\rm E})=0.\label{2s}
 \end{eqnarray}
 [Note that the last equation includes twice the
 number of solutions as Eqs. (\ref{v-k}) and (\ref{u}).]
 Here, $G_{\rm ee,oo}$ are  noninteracting,
 two-particle Green's functions, which consist of the
 even and odd (with respect to interchanging the dots)
solutions of the single-particle spectrum, respectively, with
 \begin{eqnarray}
 G_{\rm ee(oo)}({\rm E})&=&2\frac{|\phi^{\rm e(o)}_{\rm B}|^{4}}
 {{\rm E}-2\epsilon^{\rm e(o)}_{\rm B}}
 +2\sum_{k}\frac{|\phi^{\rm e(o)}_{\rm B}|^{2}
 |\phi^{\rm e(o)}_{k}|^{2}}
 {{\rm E}-\epsilon^{\rm e(o)}_{\rm B}-\epsilon_{k}}
 \nonumber\\
 &+&\frac{1}{2}\sum_{kk'}\frac{|\phi^{e(o)}_{k}|^{2}|\phi^{\rm e(o)}_{k'}|^{2}}
 {{\rm E}-\epsilon_{k}-\epsilon_{k'}}.\label{geeoo}
 \end{eqnarray}


 We next determine the ground-state energy,
 starting with the triplet states,
 whose energies are given by Eq. (\ref{v-k}).
 We assume that we are in the lower region II of
 Fig. \ref{fig1}, where there are two
 bound states below the band. (Note that
 in the regions marked 0 and I in Fig. \ref{fig1},
 where there is at most one single-particle
 bound state, the triplet `bound' state will always
 lie in the continuum.)
 The function $G_{\rm eo}({\rm E})$, Eq. (\ref{geo}),
 has the following behavior.\cite{we1,we2}
 As E approaches $-\infty $, it goes to zero
 from below. At
 E$=\epsilon^{\rm e}_{\rm B}+\epsilon^{\rm o}_{\rm B}$,
 it diverges to $-\infty$, jumps
 to $+\infty $ as E crosses that value, and then
 decreases, as E approaches the
 bottom of the
 two-electron continuum states, located at
 $-2+\epsilon^{\rm e}_{\rm B}$. As discussed in our earlier work
\cite{we1,we2}, in the thermodynamic limit of infinite `leads'
$G_{\rm eo}$ has a finite value at this band threshold,
due to the $k$-dependence of $\phi^{\rm e(o)}_{k}$ at the impurities.
The triplet
 ground-state energy is where
$G_{\rm eo}$ crosses $({\rm V-K)}^{-1}$.
 Hence, there will be a two-electron bound state only when
 $({\rm V-K})^{-1}<G_{\rm eo}(-2+\epsilon^{\rm e}_{\rm B})$.
 It follows
 that there are values of the direct and exchange Coulomb
 couplings such that the ground triplet
 state is not bound, but lies in the continuum.

 We now turn to the singlet states, again assuming the existence of
 both $\epsilon^{\rm e}_{\rm B}$ and $\epsilon^{\rm o}_{\rm B}$.
 Consider first the solutions given by Eq. (\ref{u}). Since
 U$>$V (and V--K), the lowest solution of this equation
 lies {\it above} the lowest solution
 of the triplet state, (which is given by the same
 function $G_{\rm eo}$). Hence, we need not
 consider anymore the states given by (\ref{u}).
 To explore the other family of singlet solutions,
 it is convenient to re-write Eq. (\ref{2s}) in the form
 \begin{eqnarray}
 &&\frac{1}{4}\Biggl (\frac{1}{G_{\rm ee}({\rm E})}+
 \frac{1}{G_{\rm oo}({\rm E})}\Biggr )=
 \frac{\rm U+V+K}{2}\nonumber\\
 &&\pm
 \sqrt{\frac{1}{16}\Biggl (\frac{1}{G_{\rm ee}({\rm E})}-\frac{1}{G_{\rm oo}({\rm E})}
 \Biggr )^{2}+\Biggl (\frac{\rm U-V-K}{2}\Biggr )^{2}}.\label{2s2}
 \end{eqnarray}
 Since the behavior of $G_{\rm ee}$ and $G_{\rm oo}$ as function of E is similar
 to that of $G_{\rm eo}$ described above, it follows from Eq. (\ref{2s2}) that
 the lowest singlet state energy obeys that equation with
the minus sign.

 \section{Singlet-triplet transitions}

In order to decide when the lowest bound state of
the two electrons is a singlet or a triplet,
we need to (i) determine for which values of the
Coulomb parameters Eqs. (\ref{v-k}) and
(\ref{2s2}) have  bound solutions; and (ii) to
compare these two solutions, when they exist.
Consider as an example the case in which there
is only the on-site Hubbard interaction,
that is, V=K=0. Then the triplet bound state
has the energy ${\rm E}_{\rm T}=
\epsilon^{\rm e}_{\rm B}+\epsilon^{\rm o}_{\rm B}$,
cf. Eq. (\ref{v-k}). The singlet energy,
E$_{\rm S}$,
in that case is given by the lowest solution of
\begin{eqnarray}
\frac{1}{\rm U}=G_{\rm ee}({\rm E})+G_{\rm oo}({\rm E}).\label{onlyu}
\end{eqnarray}
Similarly to the behavior of $G_{\rm eo}({\rm E})$, the
right-hand-side of this equation
starts at very small negative values
when E tends to $-\infty$. It then diverges to
$-\infty$ as E approaches $2\epsilon^{\rm e}_{\rm B}$,
jumps to $+\infty$ as E crosses that value,
diverges again to $-\infty $ as
E$\rightarrow 2\epsilon^{\rm o}_{\rm B}$,
then jumps to $+\infty$, and finally decreases towards a finite value
as E approaches the
bottom of the two-electron continuum. It
follows that Eq. (\ref{onlyu}) has always a bound energy solution.
Moreover, if $G_{\rm ee}+G_{\rm oo}$ is
{\it negative} at E$={\rm E}_{\rm T}\equiv \epsilon^{\rm e}_{\rm B}
+\epsilon^{\rm o}_{\rm B}$, then that solution
E$_{\rm S}$ lies below E$_{\rm T}$, i.e., the ground-state is a singlet.
This is indeed the case, as is shown in the Appendix [Eq. (\ref{neg})].
This is in accordance with the general
rule, which states that in order for the
ground-state to be a triplet, the exchange Coulomb energy has to overcome
the kinetic energy.

For the sake of clarity of the presentation, we will carry the
rest of the analysis to lowest order in the coupling to the leads,
$\gamma $. In that case, it is possible to derive simple
expressions for the two-particle Green's functions, see Eqs.
(\ref{sg}). Using those equations, we find that
the singlet energies are given by
\begin{eqnarray}
&&{\rm E} -2\epsilon_{0}+\gamma (e^{-\alpha^{\rm e}}+e^{-\alpha^{\rm o}})=
\frac{\rm U+V+K}{2}\nonumber\\
&-&\sqrt{
4t_{D}^{2}+\Bigl (\frac{\rm U-V-K}{2}\Bigr )^{2}
+4t_{D}\gamma (e^{-\alpha^{\rm e}}-e^{-\alpha^{\rm o}})},
\label{ES}
\end{eqnarray}
and the triplet energies are given by
\begin{eqnarray}
{\rm E}-2\epsilon_{0} +\gamma (e^{-\alpha^{\rm e}}+e^{-\alpha^{\rm o}})={\rm V-K},
\label{tt}
\end{eqnarray}
where $\alpha^{\rm e,o}$ is related to the corresponding ${\rm E}$
(with indices S or T) via
\begin{eqnarray}
{\rm E}-\epsilon^{\rm e,o}_{\rm B}=-2{\rm cosh}\alpha^{\rm e,o}.
\label{al}
\end{eqnarray}
 Let us examine the case in which in the absence of the coupling to the leads,
 the singlet and the triplet ground-state energies are equal, i.e.,
 $4t_{D}^{2}=2$K(U--V+K), see Eq. (\ref{st0}). Then
 \begin{eqnarray}
 {\rm E}_{\rm S}-{\rm E}_{\rm T}&=&\gamma (e^{-\alpha^{\rm e}_{\rm T}}
 +e^{-\alpha^{\rm o}_{\rm T}}-e^{-\alpha^{\rm e}_{\rm S}}
 -e^{-\alpha^{\rm o}_{\rm S}})
 \nonumber\\
 &-&
 \frac{4t_{D}}{{\rm U-V+3K}}\gamma (e^{-\alpha^{\rm e}_{\rm S}}
 -e^{-\alpha^{\rm o}_{\rm S}}).
 \end{eqnarray}
 The last term on the right-hand-side of this equation
 is negative. This follows from Eq. (\ref{al})
 and the fact that $\epsilon^{\rm e}_{\rm B}<\epsilon^{\rm o}_{\rm B}$.
 As for the first term, we
 use again Eq. (\ref{al}), to write it in the form
 $({\rm E}_{\rm T}-{\rm E}_{\rm S})\gamma
 ((e^{2\alpha^{\rm e}_{\rm S}} -1)^{-1}+(
 e^{2\alpha^{\rm o}_{\rm S}}-1)^{-1})$.
 Hence, E$_{\rm S}<{\rm E}_{\rm T}$ and the singlet is prefered.

\begin{figure}
 \epsfclipon
\epsfxsize = 8cm
\centerline{
\vbox{
\epsffile{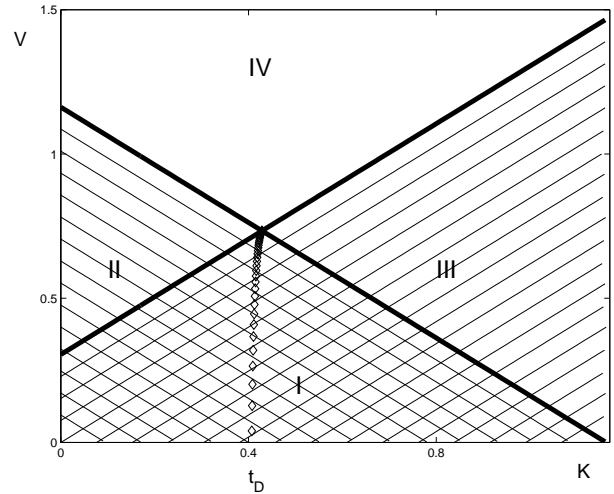}
}}
\vskip  1truecm
\caption{The phase diagram of the two-electron bound states, for
$t_{D}=0.4$, $\epsilon_{0}=-2.6$, and $\gamma =0.1$, see text.}
\label{fig2}
 \end{figure}

 The above discussion shows that the coupling
 to the continuum of propagating states
 enhances the tendency of the two electrons to
 form a singlet state, in the situation where
 in the absence of that coupling, the singlet
 and the triplet states are degenerate.  In order
 to investigate whether this tendency persists
 for other choices of parameters (and at the
 same time to keep the calculations tractable)
 we will now confine ourselves to the
 choice U=V+K.  In this case, again to leading order in $\gamma$,
the equation for
 the singlet energies (\ref{ES}) reads
 \begin{eqnarray}
 f_{\rm S}({\rm E})&=&{\rm V+K},\nonumber\\
 f_{\rm S}({\rm E})&=&{\rm E}-2\epsilon_{0}+2t_{D}
 +2\gamma e^{-\alpha^{\rm e}}.
 \label{ss}
 \end{eqnarray}
 Similarly, Eq. (\ref{tt})
 for the triplet energies
 can be written as
 \begin{eqnarray}
 f_{\rm T}({\rm E})&=&{\rm V-K},\nonumber\\
 f_{\rm T}({\rm E})&=&{\rm E}-2\epsilon_{0}+\gamma (e^{-\alpha^{\rm e}}+
 e^{-\alpha^{o}}).
 \label{ttt}
 \end{eqnarray}
 Let us first determine for which parameters
 these equations yield bound,
 two-electron energies. To this end, we consider
 Eqs. (\ref{ss}) and (\ref{ttt})
 at the bottom of the two-electron continuum,
 E=$-2+\epsilon^{\rm e}_{\rm B}$.
 The first of these equations will have a bound state for
 $f_{\rm S}(-2+\epsilon^{\rm e}_{\rm B})>{\rm V+K}$; the second will have such
 a solution when $f_{\rm T}(-2+\epsilon^{\rm e}_{\rm B})>{\rm V-K}$.
 These conditions
 are plotted in Fig. \ref{fig2} as the thick lines there.
 A bound triplet state exists in regions I+III, below the
heavy line of positive slope. A singlet bound state exists in
I+II, below the heavy line of negative slope. In region IV, there
are no bound states; both the triplet and the singlet states are
in the continuum, and their energy is about the same. Crossing the line
between regions II and IV (or III and IV) thus corresponds to the
delocalization transition discussed above, from a singlet (triplet)
bound ground state to a degenerate `metallic' state.
This transition is the most striking effect of the coupling to the leads, and
we expect it to appear irrespective of the quantitative approximation
used in Fig. \ref{fig2}.

In region I
one has to compare the singlet energy with the triplet one. These
two become degenerate along the diamond curve, whose equation is
derived from (\ref{al}), (\ref{ss}), and (\ref{ttt})
\begin{eqnarray}
{\rm V-K}&=&-\frac{\epsilon^{\rm o}_{\rm B}}{2}({\rm B}-1)
+\frac{\epsilon^{\rm e}_{\rm B}}{2}({\rm B}+1)-2\epsilon_{0}
-(1-\gamma ){\rm AB},\nonumber\\
{\rm A}&=&\frac{2}{\gamma}({\rm K}-t_{D}),\nonumber\\
{\rm B}&=&\sqrt{1+4/[{\rm A}({\rm A}+\epsilon^{\rm o}_{\rm B}-
\epsilon^{\rm e}_{\rm B})]}.
\end{eqnarray}
This line is almost vertical, with ${\rm E}_{\rm S}<{\rm E}_{\rm
T}$ to its left, and ${\rm E}_{\rm S}>{\rm E}_{\rm T}$ to its
right. The conclusion is that, as long as there exists a bound
state of the two electrons, then this line moves slightly to the right
as $\gamma$ is increased from zero
(when the line was at K$=t_D$).
However, the coupling to the
continuum states delocalizes the electrons, making the two states
degenerate over a large part of the parameter plane $\{{\rm
V-K}\}$, i. e. region IV in Fig. \ref{fig2}.
With all other parameters fixed, one might expect that increasing the
distance between the dots causes a decrease in $t_D$, in V and in K,
thus causing a shift towards to lower left side of Fig.
\ref{fig2}, towards a bound singlet ground state.

 \section{discussion}

We have derived analytical expressions for the spectrum of two
interacting electrons on a simplified model for a double
quantum-dot. When the dot is decoupled from the external leads, it
is straightforward to obtain this spectrum, and discuss the
criterion for its ground-state to be a singlet or a triplet. The
question we have addressed is how this criterion is modified when
the single-particle states which are localized on the dot are
coupled to the continuum of extended states on the leads. A typical
example of our results is shown in Fig. \ref{fig2}:
as long as the the electron-electron interactions do
not delocalize the ground-state, then the location for the singlet-triplet
transition shifts continuously
with the coupling to the leads, $\gamma$.
In that case, one can still say that
the singlet state is the ground one provided that
the kinetic energy dominates over the exchange energy. In a way,
the coupling to the leads enhances the kinetic energy, and
therefore it slightly favors the singlet ground state.
However, the coupling to the leads has a much more drastic effect:
over a significant part of the parameter space
(which
consists of the Coulomb couplings, the single-particle energies on
the dots, and the coupling to the leads), e. g. region IV in Fig.
\ref{fig2}, the interplay between the
coupling to the leads and the electron-electron
interactions delocalizes one or both electrons. Then, the
singlet and the triplet states are degenerate. In such a
situation, the bound state (which may be either a singlet or a triplet)
disappears and the ground state becomes a degenerate singlet-triplet
`metallic' state.
We believe that this delocalization effect should be taken into
consideration in the analyzes of experimental data related to this
question.

 \acknowledgements
 We enjoyed many discussions with Yoseph Imry.
This project was supported by grants from the Israel Science Foundation and
 from the Israeli Ministry of Science and the French Ministry of Research and
 Technology, AFIRST.

\appendix
\section{The two-electron Green's functions}

We first list  the eigenfunctions required for the calculation of the noninteracting,
two-particle Green's functions.  These are needed only on the double quantum
dot, that is, on the sites $r$ and $\ell $. Writing the even and odd bound-state energies,
Eqs. (\ref{eb}), in the form
\begin{eqnarray}
\epsilon_{\rm B}^{\rm e,o}=-2{\rm cosh}\kappa^{\rm e,o},
\end{eqnarray}
the wave functions on the dot sites is
\begin{eqnarray}
|\phi^{\rm e,o}_{\rm B}|^{2}=\frac{1}{2}\Biggl (1+\frac{\gamma}
{e^{2\kappa^{\rm e,o}}-1}\Biggr )^{-1}.
\end{eqnarray}
The band states have been calculated assuming periodic boundary conditions for a system of
N sites (these include the leads). Then
\begin{eqnarray}
|\phi_{k}^{\rm e,o}|^{2}=\frac{2}{{\rm N}}
\frac{\gamma\sin^{2}k}{\gamma^{2}\sin^{2}k +(\epsilon_{k}-\epsilon^{\rm e,o}_{0}
+\gamma\cos k)^{2}},\label{phik}
\end{eqnarray}
with $\epsilon_{k}=-2\cos k$, for both the even and odd states.
In calculating the sums over
$k$ in the two-particle Green's functions, we shall use the
continuum limit, dividing the
$k$-integrations on the even (odd) functions by 2.

 We next derive $G_{\rm ee}$, $G_{\rm oo}$, and $G_{\rm eo}$.
 It is convenient first to calculate the function
 \begin{eqnarray}
 Q^{\rm e,o}(\omega )=\sum_{k}\frac{|\phi^{\rm e,o}_{k}|^{2}}{\omega -\epsilon_{k}},
 \ \ \ \omega <-2.
 \end{eqnarray}
 Writing
 \begin{eqnarray}
 \omega =-2{\rm cosh}\alpha ,
 \end{eqnarray}
 and using Eq. (\ref{phik}), we find
 \begin{eqnarray}
 Q^{\rm e,o}&=&-\gamma \frac{e^{\kappa^{\rm e,o}}}{e^{2\kappa^{\rm e,o}}-1+\gamma}
 \nonumber\\
 &&\times
 \frac{e^{\alpha +\kappa^{\rm e,o}}}{(e^{\alpha +\kappa^{\rm e,o}}-1+\gamma )(e^{\alpha +
 \kappa^{\rm e,o}}-1)},
 \end{eqnarray}
 where we have also used
 \begin{eqnarray}
 \sum_{k}|\phi^{\rm e,o}_{k}|^{2}=1-2|\phi^{\rm e,o}_{\rm B}|^{2}.
 \end{eqnarray}
 Exploiting this result, we now introduce the function $F$,
 \begin{eqnarray}
 F^{\rm e,o}(\omega )&=&|\phi^{\rm e,o}_{\rm B}|^{2}+
 (\omega -\epsilon^{\rm e,o}_{\rm B})\frac{1}{2}Q^{\rm e,o}(\omega )\nonumber\\
 &=&\frac{1}{2}\times\frac{1}{1+(\gamma /(e^{\alpha +\kappa^{\rm e,o}}-1))}.\label{F}
 \end{eqnarray}
 Note that
 \begin{eqnarray}
 F^{\rm e}(\epsilon^{\rm e}_{\rm B})&=&|\phi^{\rm e}_{\rm B}|^{2},\ \ \
 F^{\rm o}(\epsilon^{\rm o}_{\rm B})=|\phi^{\rm o}_{\rm B}|^{2},\nonumber\\
 F^{\rm e}(\epsilon^{\rm o}_{\rm B})=F^{\rm o}(\epsilon^{\rm e}_{\rm B})&=&
 \frac{1}{2}\times \frac{1}{1+(\gamma /(e^{\kappa^{\rm e}+\kappa^{\rm o}}-1))}.
 \label{Fs}
 \end{eqnarray}
 It is now straightforward to show, using Eqs. (\ref{geeoo}), (\ref{geo}), and (\ref{F})
 that
 \begin{eqnarray}
 &&G_{\rm ee}({\rm E})=2\frac{(F^{\rm e}({\rm E}-\epsilon^{\rm e}_{\rm B}))^{2}}
 {{\rm E}-2\epsilon^{\rm e}_{\rm B}}\nonumber\\
 &+&\sum_{kk'}
 \frac{|\phi^{\rm e}_{k}|^{2}|\phi^{\rm e}_{k'}|^{2}
 (\epsilon^{\rm e}_{\rm B}-\epsilon_{k})(\epsilon^{\rm e}_{\rm B}-\epsilon_{k'})/2}
 {({\rm E}-\epsilon_{k}-\epsilon_{k'})
 ({\rm E}-\epsilon^{\rm e}_{\rm B}-\epsilon_{k})
 ({\rm E}-\epsilon^{\rm e}_{\rm B}-\epsilon_{k'})},\label{gee}
 \end{eqnarray}
 with an analogous result for $G_{\rm oo}$, with e replaced by o, and
 \begin{eqnarray}
 &&G_{\rm eo}({\rm E})=4\frac{F^{\rm e}({\rm E}-\epsilon^{\rm o}_{\rm B})F^{\rm o}
 ({\rm E}-\epsilon^{\rm e}_{\rm B})}{{\rm E}-
 \epsilon^{\rm e}_{\rm B}-\epsilon^{\rm o}_{\rm B}}
 \nonumber\\
 &+&\sum_{kk'}
 \frac{|\phi^{\rm o}_{k}|^{2}|\phi^{\rm e}_{k'}|^{2}
 (\epsilon^{\rm o}_{\rm B}-\epsilon_{k})(\epsilon^{\rm e}_{\rm B}-\epsilon_{k'})}
 {({\rm E}-\epsilon_{k}-\epsilon_{k'})
 ({\rm E}-\epsilon^{\rm e}_{\rm B}-\epsilon_{k})
 ({\rm E}-\epsilon^{\rm o}_{\rm B}-\epsilon_{k'})}.\label{Geo}
 \end{eqnarray}
 For energies E below the bottom of the two-electron continuum,
 i.e., E$<-2+\epsilon^{\rm e}_{\rm B}$,
 the double sum on $k$ and $k'$ in (\ref{gee}) is negative.
 Using Eq. (\ref{Fs}) for the first terms in the
 equations for $G_{\rm ee}$ and $G_{\rm oo}$, it follows that
 \begin{eqnarray}
 G_{\rm ee}(\epsilon^{\rm e}_{\rm B}
 +\epsilon^{\rm o}_{\rm B})+G_{\rm oo}(\epsilon^{\rm e}_{\rm B}
 +\epsilon^{\rm o}_{\rm B})<0.\label{neg}
 \end{eqnarray}
 This result is used to show that when the only Coulomb coupling is the Hubbard U,
 the singlet is always the ground state.

 Up to this point, the results were given for general $\gamma$.
To lowest order in the coupling to the leads, $\gamma $,
 we may discard the double sums in (\ref{gee})
 and (\ref{Geo}). Then, using (\ref{F}), we find
 \begin{eqnarray}
 \frac{1}{G_{\rm eo}({\rm E})}&&\sim {\rm E}-2\epsilon_{0}
 +\gamma (e^{-\alpha^{\rm e}}+e^{-\alpha^{\rm o}}),\nonumber\\
 \frac{1}{2G_{\rm ee}({\rm E})}&&\sim {\rm E}-2\epsilon_{0}+2t_{D}
 +2\gamma e^{-\alpha^{\rm e}},\nonumber\\
 \frac{1}{2G_{\rm oo}({\rm E})}&&\sim {\rm E}-2\epsilon_{0}-2t_{D}
 +2\gamma e^{-\alpha^{\rm o}},\label{sg}
 \end{eqnarray}
 where we have used Eq. (\ref{al}).

\end{multicols}

\end{document}